\documentclass[11pt]{article}
\usepackage{amscd}
\usepackage{verbatim}
\usepackage{amssymb}

\newcommand\nonono[1]{{}}

\begin{document}

\vskip 0 true cm \flushbottom
\begin{center}
\vspace{24pt} { \large \bf Some Applications of Ricci Flow
in Physics} \\

\vspace{30pt} {\bf E Woolgar}\footnote{ewoolgar@math.ualberta.ca} 

\vspace{24pt} 
{\footnotesize
Dept of Mathematical and Statistical Sciences, University of Alberta,\\
Edmonton, AB, Canada T6G 2G1. }\\
\vspace{24pt}
{\em Dedicated to Rafael D Sorkin on the occasion of his 60th birthday.}\\
\bigskip
Based on a keynote talk given at the Theory Canada III conference,
Edmonton, June 2007.
\end{center}
\date{\today}
\bigskip

\begin{center}
{\bf Abstract}
\end{center}


\noindent I discuss certain applications of the Ricci flow in
physics. I first review how it arises in the renormalization group
(RG) flow of a nonlinear sigma model. I then review the concept of a
Ricci soliton and recall how a soliton was used to discuss the RG
flow of mass in 2-dimensions. I then present recent results obtained
with Oliynyk on the flow of mass in higher dimensions. The final
section discusses how Ricci flow may arise in general relativity,
particularly for static metrics.
\newpage

\section{Introduction}
\setcounter{equation}{0}

\noindent The Ricci flow
\begin{equation}
\frac{\partial g_{ij}}{\partial \lambda} = -2R_{ij}\label{eq1.1}
\end{equation}
was introduced by Hamilton \cite{Hamilton} over 25 years ago. During
the intervening time, it has been studied by mathematicians
primarily as a tool for proving the geometrization conjecture for
closed 3-manifolds, which includes as a special case the Poincar\'e
conjecture that every closed, simply connected
%
%
3-manifold is a 3-sphere. This endeavor has now met with success
\cite{Perelman, MT, CZ}.

The equation describes the deformation of a Riemannian metric
$g_{ij}$ with ``time'' $\lambda$. The deformation is driven by Ricci
curvature, so that parts of the manifold with greater Ricci
curvature will undergo greater deformation. Regions with very little
Ricci curvature will change only a little during the flow. Fixed
points of the flow are the Ricci flat manifolds
\begin{equation}
R_{ij}=0 \ . \label{eq1.2}
\end{equation}

At about the time of Hamilton's work, equation (\ref{eq1.1}) was
already making its first appearance in physics in Friedan's 1980
thesis \cite{Friedan} on the renormalization group flow of nonlinear
sigma models. Equation (\ref{eq1.1}), with the coefficient $2$ on
the right-hand side replaced by a positive constant written as $T$
by Friedan but now denoted by $\alpha'$, was the renormalization
group (RG) flow equation with the $\beta$-function truncated to
leading order in $\alpha'$.

The mathematical study of the Ricci flow has experienced tremendous
growth in the aftermath of Perelman's breakthrough work
\cite{Perelman} with its consequences for geometrization.
It is a good time to look for physical applications. There are
several
possibilities, some but not all related to renormalization group
flow. A brief list of Ricci flow questions and their possible
physical applications would include:

\begin{itemize}

\item Are there classes of Lorentzian metrics to which Ricci
    flow can be applied? (I say `'classes'' in part because the
    unrestricted initial value problem would not be well-posed.)
    If so, this would surely find application in general
    relativity, but what precisely are the useful applications?

\item Is flat space stable under Ricci flow? If not, then mass
    would behave in unexpected ways under RG flow. This would
    have applications to tachyon condensation in string theory.

\item Can we do Ricci flow on manifolds with boundary? There is
    recent work on this in the context of an application to
    black hole thermodynamics \cite{HW}. As well, this question
    has some relevance to certain formulations of quasilocal
    mass.

\item Can Perelman's entropy definition be generalized to full
    RG flow (of nonlinear sigma models) to express this as a
    gradient flow on a space of coupling constants equipped with
    a positive definite metric? If so, then this would yield a
    $C$-function for this flow. See \cite{Tseytlin, OSW}.

\end{itemize}

This list is not intended to be exhaustive by any means, but is
already too large to be addressed in any reasonably complete fashion
in the present article. I will confine myself to an introductory
treatment of Ricci flow for physicists working in general relativity
and quantum field theory, together with a report on some recent
progress in two areas: the flow of mass and the flow of static
metrics.

Section II has a brief discussion of Ricci flow as an approximation
to RG flow for Sigma models. Section III introduces the important
notion of self-similar Ricci flow solutions called Ricci solitons.
Section IV reviews a discussion in \cite{GHMS}, which uses a
2-dimensional Ricci soliton to illustrate the behaviour of mass
under Ricci flow. Section V describes recent results that Oliynyk
and I have obtained regarding the mass question in dimension 3 and
greater, in spherical symmetry \cite{OW}. Section VI considers the
Ricci flow of static spacetime metrics and describes a soliton
recently found by Akbar and me \cite{AW} which can be interpreted as
generating a nontrivial Lorentzian flow.

\bigskip \noindent{\it Acknowledgements.} This work was partially
supported by an NSERC Discovery Grant to the author, who wishes to
thank the organizers for the opportunity to speak about this work at
Theory Canada III. I thank my collaborators M Akbar, T Oliynyk, and
V Suneeta for discussions and for permitting me to draw on our joint
work herein.

I first came to appreciate the importance for physics of a firm
mathematical foundation from Rafael Sorkin. Early on, together with
Penrose, we formulated an argument for a positive mass theorem which
drew on causal structure of Lorentzian manifolds but also relied
heavily on the kinds of arguments that arise in both comparison
geometry and differential topology. What I learned then has
influenced me ever since, as will I hope be at least somewhat
evident in what follows below. It is a pleasure to dedicate this
article to Rafael on the occasion of his 60th birthday.

\section{Ricci flow and Sigma Models}
\setcounter{equation}{0}

\noindent Let us start with the connection between Ricci flow and RG
flow for string theory or, more precisely, for bosonic nonlinear
sigma models from 2-dimensional spacetime to a curved Riemannian
target manifold. Consider the sigma model with only the coupling to
the target space metric $g_{ij}$ (it's easy to add a dilaton). The
action is
\begin{equation}
S=\alpha' \int_M h^{ab}(\sigma) g_{ij}(X) \frac{\partial
X^i}{\partial \sigma^a} \frac{\partial X^j}{\partial \sigma^b}
\sqrt{h} d^2\sigma \label{eq2.1}
\end{equation}
plus a boundary term if the manifold $M$ is not closed. Here
$\sigma=(\sigma^1,\sigma^2)$ denotes coordinates on the worldsheet,
$h_{ab}$ is any metric on that worldsheet, $X^i$ denotes the target
manifold coordinates of the embedded (or immersed) worldsheet, and
$\alpha'$ is a square distance (the string scale). Thinking of this
as the starting point for a theory of quantum scalar fields $X^i$ in
2 dimensions, then $g_{ij}(X)$ represents the coupling constants of
the theory and can be expected to renormalize. The renormalization
group flow can be expressed perturbatively as a power series in
$\alpha'$ \cite{Friedan}:
\begin{equation}
\frac{\partial g_{ij}}{\partial \lambda} = -\alpha' R_{ij}
-\frac{\alpha'^2}{2} R_{iklm}R_j{}^{klm}+\dots\ , \label{eq2.2}
\end{equation}
where the ellipsis denotes a power series with terms of order
$\alpha'^3$ and higher.

Now say that $g(\lambda,\cdot)$ is a solution of this equation on a
``time'' (that is, energy) interval $\lambda\in (T_1,T_2)$. Define
\begin{eqnarray}
{\hat \lambda}&:=&a\lambda \in ({\hat T}_1,{\hat T}_2):=(aT_1,aT_2)
\ , \nonumber\\
{\hat g}_{ij}({\hat \lambda},\cdot)&:=&ag_{ij}({\hat \lambda}/a,\cdot)
\equiv ag_{ij}(\lambda,\cdot)\ , \label{eq2.3}
\end{eqnarray}
for some constant $a>0$. The RG flow equation becomes
\begin{equation}
\frac{\partial {\hat g}_{ij}}{\partial {\hat \lambda}} = -\alpha'
{\hat R}_{ij} - a \frac{\alpha'^2}{2} {\hat R}_{iklm}{\hat
R}_j{}^{klm}+\dots \ . \label{eq2.4}
\end{equation}
The terms represented by the ellipsis have coefficients
$a^{p-1}\alpha'^p$ with $p\ge 3$. Thus, if we take $a\to 0$, we
obtain the Ricci flow equation (\ref{eq1.1}).

However, this does not necessarily mean that {\it solutions} of
(\ref{eq1.1}) and (\ref{eq2.4}) approach each other on arbitrary
$\lambda$ scales as $a$ becomes small. Even worse, if the original
solution $g(\lambda,p)$ was defined on a bounded interval
$\lambda\in(T_1,T_2)$, then rescaling shrinks this interval to
${\hat \lambda} \in(aT_1,aT_2)$, which tends to zero width. To avoid
this, we will primarily consider solutions $g$ defined on a
semi-infinite or infinite domain, so for any finite $a$ the domain
of ${\hat \lambda}$ is also semi-infinite or infinite. Flows on such
domains are classified as either {\it ancient} ($\lambda\in
(-\infty,0)$), {\it immortal} ($\lambda\in (0,\infty)$), or {\it
eternal} ($\lambda\in (-\infty,\infty)$). The ancient and eternal
flows are of greatest interest. If an ancient or eternal Ricci flow
stays close to a full RG flow as $\lambda\to -\infty$, then the
sigma model is well-defined when the energy cutoff is
removed.\footnote
{Think of a quantum field theory characterized by $k$ coupling
constants $c_i$, all of whose UV divergences can be traced back to
just one particular Feynman graph. Impose a momentum cutoff
$\Lambda$, so the graph takes a finite value
$\Gamma(\Lambda,c_1,\dots,c_k)=C=const$. The predictions of the
theory will now depend on the cut-off $\Lambda$ as well as on the
coupling constants $c_i$. Now follow a curve on the level set
$\Gamma=C$, along which the coupling constants $c_i$ and the cut-off
$\Lambda$ will vary but of course $\Gamma$ will not. This is an RG
flow. If one can follow some such curve until $\Lambda\to\infty$,
introducing no new divergences in other graphs, this is cutoff
removal. The limiting theory (defined by coupling constants
$c_i(\infty)$) obviously has no $\Lambda$ dependence and its Feynman
diagrams have no UV divergences. This can work even when more than
one Feynman graph is responsible for the divergences of the theory,
provided there are sufficiently many coupling constants.}

By theorem (9.15) of \cite{CK}, if an ancient or eternal solution
${\hat g}_{ij} ({\hat t})$ is asymptotically flat at each ${\hat t}$
(so the sectional curvatures are always bounded), then the scalar
curvature obeys $R\ge 0$ for all ${\hat t}$. That means the positive
mass theorem is obeyed all along the flow, and so we learn that {\em
positive target manifold mass is a requirement for a well-defined
sigma model at high energy}.

This also provokes the question of how mass behaves under RG flow or
at least, in the current context, under Ricci flow. This leads us
first to a discussion of a special solution of the flow known as a
Ricci soliton, and then to the question of whether flat space is
asymptotically stable with respect to Ricci flow.

\section{Solitons}
\setcounter{equation}{0}

\noindent As a model for Ricci flow, consider the one-dimensional
linear heat equation
\begin{equation}
\frac{\partial u}{\partial t} = \frac{\partial^2 u}{\partial x^2}
\label{eq3.1}
\end{equation}
for the temperature $u(t,x)$.

For the moment, let $x$ take values in $[0,\pi]$ and impose periodic
boundary conditions $u(t,0)=u(t,\pi)=0$. Then there are separable
solutions
\begin{equation}
u_n(t,x)=T_n(t)X_n(x)=e^{-n^2t}\sin nx\label{eq3.2}
\end{equation}
for $n$ a positive integer. Since the equation is linear, then
arbitrary linear combinations
\begin{equation}
u(t,x)=\sum\limits_{n=1}^{\infty}c_n e^{-n^2t}\sin nx\label{eq3.3}
\end{equation}
are also solutions, where the $c_n$ are constants. An elementary
argument using the maximum principle shows that this is the general
solution. The Fourier decomposition of the initial data $u(0,x)$
uniquely determines the constants $c_n$.

Separable solutions (\ref{eq3.2}) are {\it self-similar}.
Self-similarity means that they evolve in time purely by rescaling
the amplitude. Now the general solution is not self-similar because
it is not separable. Self-similar solutions arise only for very
special initial data. Specifically, if the initial data contain
precisely one Fourier mode, say the $n^{\rm th}$ mode, then the
particular solution evolving from that data is $u_n(t,x)=c_n
e^{-n^2t}\sin nx$ and is self-similar and separable.

If the initial data contain many Fourier modes, the lowest of which
is the $n^{\rm th}$ mode, then at late times the higher modes $m>n$
are suppressed relative to this mode by a factor of
$e^{-(m^2-n^2)t}$, so the solution tends to $e^{-n^2t}\sin nx$.
Solutions always approach self-similarity at late times.

In Ricci flow, self-similar solutions exist and are called {\it
Ricci solitons}. They are always either anceint, immortal, or
eternal, and so are of interest for us. Of course, in contrast to
the heat equation case, an arbitrary Ricci flow is not a sum of
solitons, since Ricci flow is nonlinear, but general flows sometimes
asymptote to solitons in certain limits.

{\it Steady} solitons do not rescale under Ricci flow. Any
Ricci-flat metric is an example of a steady soliton. There are also
{\it expanding} and {\it shrinking} solitons, which do rescale. A
simple example of a shrinking soliton is the round shrinking
$n$-sphere
\begin{equation}
g_{ij}(\lambda)=2(n-1)(\Lambda-\lambda)g_{ij}^{\rm can} \ , \
\Lambda=const\ , \label{eq3.4}
\end{equation}
where $g^{\rm can}$ denotes the canonical sphere metric
$d\theta_1^2+\sin^2\theta_1 \left ( d\theta_2^2 + \dots \right )$.
This solution exists for$\lambda\in(-\infty,\Lambda)$ and is said to
``extinguish'' at $\lambda=\Lambda$. Like the lowest Fourier mode in
heat flow, this soliton ``attracts'' in the sense that any slightly
deformed sphere will approach the round shrinking sphere as flows
toward extinction.

A more sophisticated example is the solvegeometry expanding soliton
(Bianchi type IV$_{-1}$)
\begin{equation}
ds^2=e^{-2z}dx^2+e^{2z}dy^2+4\lambda dz^2\ . \label{eq3.5}
\end{equation}
This metric has a discrete group of isometries such that the
quotient space is a smooth compact manifold \cite{Thurston}, but the
soliton interpretation is valid only on $\Re^3$, not on a quotient.

Why is this a Ricci soliton? To answer, we can directly compute the
Ricci curvature of (\ref{eq3.5}), and we get $R_{ij}= -2 \delta^z_i
\delta^z_j$. Also by direct computation, we see that $\frac{\partial
g_{ij}}{\partial \lambda}= 4\delta^z_i\delta^z_j$. Comparing these
two results, we see that (\ref{eq1.1}) is satisfied. So we have a
Ricci flow, but is it self-similar? To answer this, we change
coordinates and write the metric (\ref{eq3.5}) as
\begin{equation}
ds^2=g_{ij}({\tilde x}) d{\tilde x}^i d{\tilde x}^j
=4\lambda \left ( e^{-2{\tilde z}}d{\tilde x}^2
+e^{2{\tilde z}}d{\tilde y}^2+d{\tilde z}^2\right )
= 4\lambda g_{ij}^0({\tilde x}) d{\tilde x}^i d{\tilde x}^j
\ , \label{eq3.6}
\end{equation}
where we have performed a $\lambda$-dependent diffeomorphism
$\phi_{\lambda}$ taking $x\mapsto {\tilde
x}=\frac{x}{\sqrt{4\lambda}}$, $y\mapsto {\tilde y} =
\frac{y}{\sqrt{4\lambda}}$, and $z\mapsto {\tilde z}=z$. Since the
coefficients $g^0_{ij}$ do not vary with $\lambda$, $ds^2$ evolves
self-similarly; i.e., it evolves only through the overall scale
factor $4\lambda$ in front of $g^0$ {\it and} via the
$\lambda$-dependent diffeomorphism $\phi_{\lambda}$.

However, now we have from (\ref{eq3.6}) that $\frac{\partial
g_{ij}}{\partial \lambda}=g^0_{ij}$ , while the Ricci curvature has
not changed, and so now $\frac{\partial g_{ij}}{\partial
\lambda}\neq -2R_{ij}$! This shows that (\ref{eq1.1}) is {\it not
geometrical}, in that it depends on the choice of coordinates along
the flow. There should really be an extra Lie derivative term
accounting for the fact that the coordinates can vary with
$\lambda$. This is a new feature not seen in our scalar
1-dimensional heat equation analogy, and gives rise to the {\it
Hamilton-DeTurck equation}
\begin{equation}
\frac{\partial g_{ij}}{\partial \lambda} = -2R_{ij}
+ \pounds_X g_{ij}\ , \label{eq3.7}
\end{equation}
where the vector field $X$ generates the diffeomorphism
$\phi_{\lambda}$ by which the coordinates vary along the flow and
$\pounds_X$ denotes the Lie derivative along $X$. This also gives
rise to the {\it Ricci soliton equation}
\begin{equation}
R_{ij} -\frac12\pounds_X g_{ij}-c g_{ij} =0\ , \label{eq3.8}
\end{equation}
where $c$ is a constant. If $g$ is a solution of this equation, then
there is a Ricci flow given by $a(\lambda)\phi^*_{\lambda}g$, and so
this flow is self-similar. For the solvegeometry, the metric $g^0$
is a solution of (\ref{eq3.8}).

\section{The Mass of a 2-Dimensional Soliton}
\setcounter{equation}{0}

\noindent We now begin our discussion of the behaviour of mass under
Ricci flow. Before passing to the asymptotically flat case, this
section will discuss the 2-dimensional, rotationally symmetric,
expanding soliton used in a study of mass under RG flow \cite{GHMS}.
This soliton is immortal but not eternal. It's a gradient soliton,
meaning that the vector field $X$ generating the soliton is the
gradient of a scalar potential $X^i=g^{ij}\nabla_j\varphi$. The
metric and potential function can be written as
\begin{eqnarray}
ds^2 &=& \lambda \left ( f^2(r) dr^2 + r^2 d\theta^2 \right )
\ , \nonumber \\
\varphi(r)&=& \int\limits^r r'f(r')dr'\ , \label{eq4.1}
\end{eqnarray}
where $f$ is given implicitly by
\begin{equation}
\left ( \frac{1}{\zeta}-1\right ) \exp \left (
\frac{1}{\zeta}-1-\frac{x^2}{2\alpha'} \right ) = \left (
\frac{1}{f(x)}-1\right ) \exp \left ( \frac{1}{f(x)}-1\right )\ ,
\label{eq4.2}
\end{equation}
where $\zeta=const$. What is important is that $f(x)\to\zeta$ for
$x\to 0^+$, $f(x)\to 1$ for $x\to \infty$, and $f$ is monotonic in
between. Changing coordinates to $\rho=r\sqrt{\lambda}$, we get
\begin{eqnarray}
ds^2 &=& f^2(\rho/\sqrt{\lambda}) d\rho^2 + \rho^2 d\theta^2 \ .
\label{eq4.3}
\end{eqnarray}

Now take $\theta\in [0,2\pi\zeta]$. For $\lambda\to 0^+$, at any
fixed $\rho>0$ (and thus at any fixed proper distance from the
origin), we have $f(\rho/\sqrt{\lambda})\to 1$. The limiting metric
(in the Gromov-Hausdorff sense: at each $\lambda$ we fix a point,
here the origin, and study the metric within arbitrarily large
proper radius balls about this point) is a flat cone of deficit
angle $\delta = 2\pi (1-\zeta)$. This is the special initial
condition which gives rise to this soliton. Now in 2 dimensions,
deficit angle plays the role of mass, so $2\pi (1-\zeta)$ is the
mass.

Since this is an expanding soliton, one might at first expect the
mass $\delta$ to rescale with $\lambda$ but this is not what
happens. At any fixed $\lambda>0$, the metric is asymptotic at large
$\rho$ to the same flat cone, so the mass remains the same. But at
any $\lambda>0$, there is no cone point (since
$f(\rho/\sqrt{\lambda}) \to \zeta$ for $\rho\to 0$ with $\lambda>0$
fixed), so the manifold is smooth.

However, as $\lambda\to\infty$, the curvature within any fixed
distance of the origin dissipates and so the Gromov-Hausdorff limit
is just flat space. Thus, if the initial $\lambda=0$ flat cone has a
positive deficit angle (i.e., if $0<\zeta<1$), the mass
``decreases'' in this sense to a flat and thus zero mass metric in
the infinite time limit; if the initial deficit angle is negative
(i.e., if $\zeta>1$), the mass increases to zero in this
sense.\footnote
{Smoothness at the origin during the flow is actually a boundary
condition imposed at the origin, where the polar coordinate system
breaks down. One can impose any deficit angle one wishes at the
origin instead. Then the $\lambda\to\infty$ limiting metric will be
a flat cone with this deficit angle. Therefore, this solution can
describe transitions between two different flat cones.}
Either way, the flow is attracted to flat space in the limit of
large $\lambda$, but at any finite $\lambda$ the mass (=deficit
angle) is unchanged from its initial value.

\section{Mass in Higher Dimensions}
\setcounter{equation}{0}

\noindent In dimension $n\ge 3$, we can define asymptotic flatness
by choosing suitable coordinates near infinity which will be
Cartesian coordinates for a flat reference metric. An asymptotically
flat metric is one whose components differ from the reference by an
element of some $H^k_{\delta}$ space, where the fall-off rate
$\delta$ (not related to $\delta$ in the previous section) is
sufficient to give a well-defined ADM mass. For defintions and
details, see \cite{OW}.

Returning momentarily to the heat equation analogy, recall that the
1-dimensional heat equation on an infinite domain has a certain
non-uniqueness. Namely, Tychonoff pointed out that
\begin{equation}
u(t,x)=\sum\limits_{k=0}^{\infty} \frac{x^{2k}}{(2k)!}
\frac{d^k}{dt^k} e^{-1/t^2} \label{eq5.1}
\end{equation}
solves the 1-dimensional heat equation for $t\in (0,\infty)$ and
$\lim_{t\to 0^+} u(t,x) =0$. Thus, we can add an arbitrary multiple
of (\ref{eq5.1}) to any solution to get a new solution for the same
initial data. An extreme case is if the temperature is initially
zero everywhere, but at any positive time the temperature
distribution can be that described by (\ref{eq5.1}), and so is large
at large $x$.

This shows that it is necessary to impose fall-off conditions on
solutions of the heat equation all along the flow, not just on the
initial temperature distribution, in order to have uniqueness.
Remarkably, this problem does not seem to arise for Ricci flow. At
least, asymptotic flatness of initial data is always preserved so
long as the flow remains nonsingular \cite{List, OW2}. Not only
that, but the mass does not change, just as we saw in the
rotationally symmetric 2-dimensional example of the last section,
and as we will now see in higher dimensions (without assuming any
symmetry).

Because asymptotic flatness is preserved, we can evaluate the mass
at any time along the flow. If the mass evaluated at two different
$\lambda$-values differs, then at some intermediate $\lambda$-value
the mass must have nonzero flow derivative. But the following simple
computation shows that this cannot happen.

On an asymptotic end $U\subseteq M$, let $\delta_{ab}$ be a fixed
(constant along the flow) Euclidean metric compatible with the
derivative $\partial_a$ and let $\Sigma$ be a closed embedded
hypersphere. Define the functional $E_{\Sigma}:S^2U\to \Re$ acting
on symmetric covariant 2-tensors by
\begin{equation}
E_{\Sigma}(g):=\frac{1}{16\pi}\int_{\Sigma}\left [ \eta^i
\partial^j g_{ij} - \eta\cdot\partial \left ( \delta^{ij}g_{ij}\right )
\right ] dA(\delta)\ , \label{eq5.2}
\end{equation}
where $\partial^i:=\delta^{ij}\partial_j$ and $\eta$ is the outward
unit normal vector to $\Sigma$ wrt the metric $\delta$. If $\Sigma$
belongs to a sequence of hyperspheres $\Sigma_n$ whose interiors
exhaust $U$, then in the limit we get the ADM mass of $g$:
\begin{equation}
E^{\rm ADM}=\lim_{n\to\infty}E_{\Sigma_n}(g)\ . \label{eq5.3}
\end{equation}
Now this functional is linear in $g$ so
\begin{equation}
\frac{dE_\Sigma}{dt}=E_\Sigma\left (\frac{\partial
g}{\partial t} \right ) =\frac{1}{16\pi}\int_{\Sigma}\left [ \eta^i
\partial^j \frac{\partial g_{ij}}{\partial t}- \eta\cdot\partial
\left ( \delta^{ij}\frac{\partial g_{ij}}{\partial t}\right )\right
]dA(\delta)\ . \label{eq5.4}
\end{equation}
If the metric evolves by Ricci flow (\ref{eq1.1}) then we get
\begin{eqnarray}
\frac{dE_\Sigma}{dt}&=&E_\Sigma\left (\frac{\partial
g}{\partial t} \right ) =-\frac{1}{8\pi}\int_{\Sigma}\left [
\eta^i\partial^j R_{ij}- \eta\cdot\partial \left (
\delta^{ij}R_{ij}\right )\right ]dA(\delta)\nonumber \\
&=&\frac{1}{8\pi}\int_{\Sigma} dA(g)\left \{ \frac{1}{2} \eta\cdot
\nabla R + {\cal O}(1/r^n) \right \} \in {\cal O}(1/r^2) \to 0\ ,
\label{eq5.5}
\end{eqnarray}
where we used the Bianchi identity to pass from one line to the next
and used asymptotic flatness to treat derivatives compatible with
$\delta$ and $g$ as equivalent near infinity (they differ by terms
that fall off fast enough to ignore). We can write the asymptotic
flatness condition in the succinct form $\partial^k (g-\delta) \in
{\cal O}(1/r^{n+k-2})$ for $k=0,1,2,3$. This gives immediately that
$\nabla R \sim \partial^3 g \sim 1/r^{n+1}$ while $dA\sim r^{n-1}$,
so the integral falls off as claimed in the last step of
(\ref{eq5.5}). Hence the derivative of the mass is always zero and
so the mass is constant, just as in the special 2-dimensional
example of the last section.

Another feature of the 2-dimensional soliton appears relevant in
higher dimensions. For rotationally symmetric metrics, the flow
converges to flat space \cite{OW}. This result assumes the initial
data for the flow are asymptotically flat and contain no minimal
hypersphere (if such a sphere were present, it might pinch off), but
the data are otherwise quite arbitrary. Such a flow is then immortal
and (Cheeger-Gromoll) converges to the flat metric in the limit of
infinite flow time. Moreover, if the initial metric is not
spherically symmetric but is in a certain sense close to being flat,
the flow is also immortal and converges to a flat metric \cite{SSS}.
(The notion of ``close to flat'' used in \cite{SSS} is too
restrictive to allow nonzero mass in 3 dimensions, but allows it in
all higher dimensions.)

It may seem paradoxical that ADM mass is constant throughout the
flow and yet the limiting metric is flat. Certainly then
$\lim_{\lambda \to \infty}E_{\rm ADM}(g(\lambda))\neq E_{\rm
ADM}(g(\infty))=0$. The resolution is that the notion of convergence
of manifolds in the $\lambda\to\infty$ limit involves picking a
marked point (e.g., the origin of symmetry in the rotationally
symmetric case) and examining the geometry of sequences of
arbitrarily large metric balls about that point. Thus, the
quasilocal mass of these balls is relevant. Oliynyk and I studied
the decay of quasilocal mass in our rotationally symmetric flow
\cite{OW} and found that within any ball of fixed (proper) radius,
the quasilocal mass evaporates at a rate $\sim 1/\lambda$. The
coefficient of this rate depends on the radius and vanishes at
infinity fast enough that the ADM mass remains constant at any
finite $\lambda$.

\section{Static Lorentzian Metrics}
\setcounter{equation}{0}

\noindent The discussion of the previous section points to the
likely importance that an understanding of the flow of quasilocal
mass would have for the general question of stability of flat space
(and other asymptotically flat fixed points) when no symmetry is
assumed. Specifically, an understanding of the conditions under
which the mass evaporates monotonically in $\lambda$ is likely to be
an important tool.

Conversely, the Ricci flow (and the closely related flow below) may
lead to better understanding of quasilocal mass. Bartnik has
proposed a definition of quasi-local mass; for details, see
\cite{Bartnik}. Say that $B$ is a Riemannian manifold\footnote
{This is the ``spatial'' version of the problem. There is also a
``spacetime'' version.}
with boundary $\partial B$. Embed $B$ in an asymptotically flat
manifold $(M,g)$ that has non-negative scalar curvature and contains
no minimal surface other than possibly a so-called compact outermost
horizon. Compute the ADM mass of $(M,g)$. The infimum of the ADM
mass over all such possible extensions $(M,g)$ is the Bartnik
quasilocal mass of $B$.

Quasilocal mass is important only insofar as it provides a {\it
useful} geometric characterization of bounded regions within a
manifold. As such, the quasilocal mass must have useful properties,
among which monotonicity (if $B_1\subseteq B_2$, then $m_{B_1}\le
m_{B_2}$) and positivity are classic examples \cite{Geroch, HI}.
Bartnik's mass has both of these. But is it nonzero in general? Yes,
if the infimum in the definition were a genuine minimum. That is, if
there were a specific asymptotically flat but nonflat $R\ge 0$
extension $(M,g)\supset B$ whose ADM mass were $\le$ that of any
other extension, then this ADM mass and, thus, the Bartnik mass of
$B$, would necessarily be $>0$ (by the positive mass theorem).

Bartnik \cite{Bartnik} conjectured that a mass-minimizing extension
exists and outside of $B$ is a solution of the static Einstein
equations. These equations are
\begin{eqnarray}
R_{ij}&=&\nabla_i u \nabla_j u\ , \label{eq6.1}\\
\Delta u &=& 0 \ , \label{eq6.2}
\end{eqnarray}
where the second equation is merely an integrability condition for
the first; it follows by applying a divergence to (\ref{eq6.1}) and
using the contracted second Bianchi identity. Given a solution
$(u,g)$ of the fixed point equations, then the metric
\begin{equation}
ds^2\equiv G_{\mu\nu}dx^{\mu}dx^{\nu}
=-e^{2u}dt^2+e^{-\frac{2u}{\sqrt{(n-1)(n-2)}}}g_{ij}dx^i dx^j
\label{eq6.3}
\end{equation}
is Ricci-flat and static.

But how does one produce a mass-minimizing metric on the extension
of $B$? One suggestion is to extend $B$ by an arbitrary
asymptotically flat metric obeying ``geometric boundary conditions''
(that the induced metric and mean curvature of $\partial B$ should
match from both sides) and having $R\ge 0$, and then to flow this
metric. The problem then is to find a flow which exists subject to
these boundary conditions, preserves $R\ge 0$ and asymptotic
flatness, and whose fixed points are solutions of (\ref{eq6.1}). If
this flow converges to a fixed point, then perhaps we would obtain
sequences of metrics converging on a possible minimizer (though if
the flow is similar to Ricci flow, we expect the ADM mass to remain
constant along the sequence and "jump" at the limit).

Such a flow was studied by List \cite{List} in his PhD dissertation
under the direction of G Huisken. The flow equations are
\begin{eqnarray}
\frac{\partial g_{ij}}{\partial \lambda} &=& -2\left (
R_{ij}-\nabla_iu\nabla_ju\right ) \ , \label{eq6.4}\\
\frac{\partial u}{\partial \lambda} &=& \Delta u\ ,
\label{eq6.5}
\end{eqnarray}
For reasonable fall-off (say $\partial u\sim 1/r$, $\partial^2 u
\sim 1/r^2$), ADM mass remains constant along this flow, just as it
does for Ricci flow.

If $(u,g)$ satisfies the flow equations (\ref{eq6.4}, \ref{eq6.5})
then the metric
\begin{equation}
ds^2\equiv g_{\mu\nu}dx^{\mu}dx^{\nu}=\pm e^{2u}dt^2+g_{ij}dx^i dx^j
\label{eq6.6}
\end{equation}
solves the Hamilton-DeTurck flow
\begin{equation}
\frac{\partial g_{\mu\nu}}{\partial \lambda} = -2R_{\mu\nu}
+ \pounds_{-\nabla u} g_{\mu\nu}\ , \label{eq6.7}
\end{equation}
with diffeomorphism generated by $-\nabla u$, where $R_{\mu\nu}$
refers to the metric $g_{\mu\nu}$ of (\ref{eq6.6}). This opens up
some possibilities. For example, if we take $g_{ij}$ to have
Lorentizian signature, then (\ref{eq6.1}) is the Einstein equation
in the presence of a free (massless) scalar field. Many solutions
are known (e.g., \cite{JNW, Wyman}). Taken together with the ``$+$''
sign in (\ref{eq6.6}), this produces a Lorentzian Ricci
soliton.\footnote
{It will be only a local soliton: the singularities of Einstein
theory will be present of course; see \cite{Chase} for a singularity
theorem tailored to this.}

If $g_{ij}$ has Euclidean signature, non-negative Ricci curvature
(as implied by (\ref{eq6.1})) is an obstruction to finding
nonconstant bounded-below (or above) solutions of (\ref{eq6.2})
\cite{Yau}. Then $u$ will be unbounded above and below, which is not
unacceptable, but a minor modification will circumvent this in any
case. Specifically, Akbar and I \cite{AW} consider instead the
Einstein-free scalar system with a cosmological term
\begin{equation}
R_{ij}=\nabla_iu\nabla_ju+\kappa g_{ij}\ , \label{eq6.8}
\end{equation}
where we will take $\kappa=const<0$ to make the Ricci curvature
nonpositive. As before, (\ref{eq6.2}) is an integrability condition
for this equation as well. Then it's straightforward (see \cite{AW})
to show that

\medskip
\noindent{\bf Lemma.} {\sl The metric (\ref{eq6.6}) constructed from
any solution $(u,g_{ij})$ of (\ref{eq6.8}) obeys
\begin{equation}
R_{\mu\nu}-\frac12\pounds_X g_{\mu\nu}-\kappa g_{\mu\nu}=0\ ,
\label{eq6.9}
\end{equation}
with $X=-\kappa t \frac{\partial}{\partial t} - g^{ij}\nabla_i u
\frac{\partial}{\partial x^j}$, and is therefore a Ricci soliton.}
\medskip

One example is given by the metric
\begin{equation}
g_{ij}dx^idx^j= dx^2+\left (1+e^{\sqrt{2}x}\right )
\left ( d\theta^2+\sinh^2\theta d\phi^2\right ) \label{eq6.10}
\end{equation}
on either $M=\Re^3$ or $M=\Re\times{\cal H}(g)$ with ${\cal H}(g)$ a
compact Riemann surface of genus $g>1$. The scalar field is
\begin{equation}
u=x-\frac{1}{\sqrt{2}}\log \left ( 1+e^{\sqrt{2}x}\right )\ .
\label{eq6.11}
\end{equation}
Clearly this metric is complete and $u$ is everywhere defined on
$M$. The corresponding soliton metric
\begin{equation}
ds^2=\pm\frac{e^{2x}}{\left ( 1+e^{\sqrt{2}x} \right )^{\sqrt{2}}}
dt^2+dx^2+\left (1+e^{\sqrt{2}x}\right )
\left ( d\theta^2+\sinh^2\theta d\phi^2\right ) \label{eq6.12}
\end{equation}
solves (\ref{eq6.9}) with $\kappa=-1$. It is complete and has
bounded sectional curvature. This produces a nontrivial immortal
Ricci flow on a nonsingular Lorentzian manifold.

On the issue of the behaviour of mass under the flow (\ref{eq6.4},
\ref{eq6.5}) (particularly, in the limit of approach to fixed
points), work continues.

\end{document}